# CARBON NANOTUBES SYNTHESIZED IN CHANNELS OF AlPO$_4$-5 SINGLE CRYSTALS : FIRST X-RAY SCATTERING INVESTIGATIONS

P. Launois[1,*], R. Moret[1], D. Le Bolloc'h[2], P.A. Albouy[1], Z.K. Tang[3], G. Li[3], J. Chen[4]

[1] *Laboratoire de Physique des Solides (UMR CNRS 8502), bât. 510, Université Paris Sud, 91405 Orsay cédex, France*
[2] *ESRF, BP 220, 38043 Grenoble cédex, France*
[3] *Physics Department, Hong Kong University of Science and Technology, Clear Water Bay, Kowloon, Hong Kong*
[4] *Chemistry Department, Jilin University, Changshun, China*

**Abstract**

Following the synthesis of aligned single-wall carbon nanotubes in the channels of AlPO$_4$-5 zeolite single crystals[1], we present the first X-ray diffraction and diffuse scattering results. They can be analysed in terms of a partial filling of the zeolite channels by nanotubes with diameter around 4Å. The possible selection of only one type of nanotube during the synthesis, due to the constraints imposed by the zeolite host, is discussed.

## 1. Introduction

Carbon nanotubes have raised a tremendous interest since their discovery in 1991[2,3]. Single wall carbon nanotubes (SWCNT) consist of a layer of graphite rolled up into a cylinder (figure 1(b)-(d)). They are highly one-dimensional materials, with typical radii around 1nm and lengths of the order of µms. The physical properties of nanotubes are strongly related to their structure[3], making structural studies of particular interest. X-ray scattering experiments should allow one to characterise macroscopic assemblies of nanotubes and to probe the internal structure of the nanotubes together with their correlations and interactions. However high quality samples (such as aligned nanotube samples) are needed to obtain detailed and reliable information. Large scale production of SWCNTs is presently achieved by laser vaporisation[4] or by arc discharge[5] techniques. The samples consist of a powder of randomly oriented nanotubes mixed with catalyst particles and with other carbon species, which limits the information that can be obtained from X-ray diffraction. Chemical vapour deposition methods also appear very promising for the synthesis of nanotubes[6]. Aligned nanotube arrays have even been realized[6(a)], but for multiwall nanotubes only. Alternatively, template techniques are appealing for obtaining aligned nanotubes. For instance, aligned multiwall nanotubes have been obtained by carbonization in anodic aluminium oxide films[7]. The synthesis of SWCNTs in the channels of AlPO$_4$-5 zeolite single crystals[1] (AFI in the zeolite terminology[8]) as templates appears to be an interesting way for obtaining aligned SWCNTs. The AFI framework contains parallel open channels[9] (figure 1(a)); the crystals are grown with tripropylamine (TPA) inside these channels[10] and the nanotubes are subsequently formed by pyrolysis of the TPA molecules in vacuum. An excess of TPA was continuously provided by introducing $5 \times 10^{-3}$ Torr pressure of TPA molecules to the growth system to increase the carbon concentration . Evidence for the presence of the nanotubes has been given by polarised

Raman scattering data[1,11,12], which exhibit a characteristic radial breathing mode. It is supported by transmission electron microscopy[1,13] and by X-ray photoelectron spectroscopy[12]. In this article, we report on the first X-ray scattering study performed on the AFI single crystals containing nanotubes.

## 2. X-ray diffraction and diffuse scattering results

$AlPO_4$-5 single crystals have elongated hexagonal prismatic shapes and their typical dimensions are 200 µm in cross-section and 500 µm in length. The samples were prepared in Hong Kong. The X-ray scattering experiments have first been performed in Orsay with conventional X-ray sources, using monochromated CuKα radiation ($\lambda$=1.5418 Å). Four crystals with nanotubes in the channels (named AFI-nanotube) have been studied. AFI crystals with empty channels (named AFI-hollow) can be obtained by high temperature treatment[12]. Two such crystals and two crystals containing TPA (named AFI-TPA) have been examined for the sake of comparison. Further experiments have been performed on one of the AFI-nanotube crystal using synchrotron radiation at the ESRF (ID01 beam line, wavelength $\lambda$=1.044 Å). Diffraction patterns were recorded on photographic films or on imaging plates for the Orsay experiments and with a 2D gas-filled wire detector at ESRF.

The AFI-nanotube crystals were characterised using the precession technique, which gives undistorted patterns of selected reciprocal planes. The lattice is found to be hexagonal with cell parameters : a=b≈13.65Å (±0.1Å), c≈8.4Å (±0.1Å) and γ=120°, which are close to those of AFI-hollow crystals[9]. Several additional peaks of weak intensity are present. They form layers normal to **c*** at wave vectors $q_{s1}$≈0.032 Å$^{-1}$ and $q_{s2}$≈0.046 Å$^{-1}$, relative to the AFI parent Bragg peaks. These satellites correspond to incommensurate periods of about 31 and 22 Å along **c**.

To probe disorder or local order via their diffuse scattering effects we performed experiments under vacuum to optimise the signal over background ratio. Diffraction patterns obtained with the AFI-hollow, AFI-TPA and AFI-nanotube crystals reveal complex diffuse scattering features which are signatures for disorder or local order phenomena related to the AFI network, the TPA molecules and the nanotubes (figure 2). The comparison of these diffuse scattering features reveals that one of them is characteristic of the AFI-nanotube crystals (it is absent in the AFI-TPA and AFI-hollow crystal patterns); it consists of a diffuse scattering sheet in the plane l=0 and at low wave vector Q. Its scattering width in the **c*** direction is similar to that of the Bragg peaks, corresponding to a few hundred Å. Its intensity does not depend on the orientation angle around the axis **c*** and thus presents a circular symmetry. The wave vector dependence of the intensity is shown in figure 3. The intensity decreases with increasing wave vector. The useful wave-vector range is upper limited to Q ~ 0.2Å$^{-1}$ because some other type of diffuse scattering has been observed above 0.2Å$^{-1}$ in AFI-TPA crystals.

## 3. Analysis and discussion

The incommensurate satellites at $q_{s1}$ and $q_{s2}$ are found to be attached to the AFI parent Bragg peaks. They can thus be attributed to an incommensurate modulation of the AFI host lattice. Further structural studies should be undertaken to determine whether this modulation is due to the interactions with nanotubes present inside the channels.

Here we focus on the l=0 diffuse scattering close to the origin. Its width in the **c*** direction being resolution limited, it corresponds to scattering by objects possessing long range order along the **c** direction, like the nanotubes. Diffuse scattering from nanotubes in the AFI channels can result from : i) orientational and/or translational disorder of the nanotubes inside the channels, or ii) incomplete filling of the channels by the nanotubes, or both. We examine these two cases below.

For uncorrelated orientational disorder between the nanotubes and for complete translational disorder along the channel axis **c**, Bragg scattering from the nanotubes only occurs in the l=0 plane, at the same reciprocal lattice nodes as those of the AFI host structure. The diffuse scattering due to the disorder writes :

$$I_D(\mathbf{Q}) \propto \left[\langle F(\mathbf{Q})F^*(\mathbf{Q})\rangle - |\langle F(\mathbf{Q})\rangle|^2\right]\delta(Q_z) + \sum_{m \neq 0}\left[\langle F(\mathbf{Q})F^*(\mathbf{Q})\rangle\right]\delta\left(Q_z - \frac{m}{T}\right) \quad \text{(Eq. 1)},$$

where **Q** is the scattering vector, $Q_z$ is its component along $\mathbf{c}^*$ and T is the nanotube period along its long axis (parallel to **c**). $F(\mathbf{Q})$ is the form factor of a nanotube unit cell of length T and the mean-value is taken over the nanotube orientations around their long axis. Delta functions assume that the nanotubes are long enough compared with the X-ray resolution. The first term in eq.1 corresponds to diffuse scattering in the plane l=0 and the second one to diffuse scattering in planes at m/T along **c***. Experimentally no diffuse scattering has been measured in the non-equatorial planes. Further experiments are needed to state whether it was too weak to be detected or if the nanotubes are well ordered. Moreover, the diffuse intensity close to the origin in the l=0 plane is calculated to be negligible. Indeed, for small wave vectors, the nanotubes can be approximated as homogeneous cylinders and the first term in eq.1 vanishes. The hypothesis of orientational and translational disorder thus fails to explain the observed scattering at small wave vectors in the l=0 plane.

On the opposite, if the channels are not completely filled by the nanotubes and if the projected electron density along the **c** direction differs from channel to channel, additional diffuse scattering is expected in the plane l=0, due to contrast density. At small wave vectors, it is proportional to the Fourier transform of a carbon cylinder, which is the zero-order Bessel function $J_0$ :

$$I_D(\mathbf{Q}) \propto f_c(Q)^2 J_0(2\pi QR)^2 \delta(Q_z) \quad \text{(Eq.2)}.$$

Q is the wave vector modulus, R the nanotube radius and $f_C$ the carbon form factor. Calculation of the diffuse intensity from eq. 2, for nanotube diameters equal to 4 and 5 Å, is plotted in figure 3. A constant background component has been included in the calculation. The experimental intensity values are rather well reproduced for nanotubes with 4Å diameters. Diameters of 5Å can be considered as an upper limit. The 4Å value is compatible with the value of 7.3Å for the zeolite channel crystallographic free diameter[8], after subtraction of twice the carbon van der Waals radius (~1.7Å). The strain energy of nanotubes with such small diameter is large[14] (nanotube diameters are usually larger that 7Å). Their stabilisation is probably to be attributed to the interactions with the zeolite host. Nanotubes of 5Å diameter have been reported very recently as the inner part of some multi-wall carbon tubes[15], where interactions with the surrounding walls may play an important role.

In brief, our X-ray results, although they are not sufficiently precise to allow us to ascertain that the proposed interpretation is unique, can be well interpreted as being due to a partial filling of the zeolite channels by nanotubes whose diameter is about 4Å. Moreover, our finding of a nanotube diameter around 4Å is in good agreement with 4.2 Å, recently observed using TEM[13]. It is also in agreement with Raman investigations[11], where a peak centred at 530cm$^{-1}$ has been assigned to the nanotube radial breathing mode. Indeed, the radial breathing mode frequency of nanotubes is strongly dependent on nanotube diameter, and an extrapolation of the dependence of the frequency versus diameter[16] gives frequencies ranging from 560 to 450 cm$^{-1}$ for nanotube diameters between 4 and 5 Å (these extrapolated frequencies are qualitative all the more so as the interactions between the nanotubes and the zeolite can modify the radial breathing mode frequency).

Finally, let us suggest an interesting hypothesis concerning possible structural relationships between the nanotubes and the zeolite. The structural characteristics of all possible nanotubes with diameters ranging between 4 and 5 Å are listed in Table 1. The period of the (5,0) and (6,0) nanotubes, T=4.3Å, is close to half the zeolite period c=8.4 Å, which is in fact the internal period of the oxygen rings forming the channels. We thus propose that interactions with the AFI during the nanotube formation may lead to the selection of only one type of nanotube ((5,0) or (6,0), fig.1(c) and (d)). Contrarily, the other methods of synthesis[4,5] lead to nanotubes with mixed diameters and helicities[17]. Further X-ray scattering experiments are planned to confirm this attractive hypothesis.

## 4. Conclusion

We have performed the first X-ray scattering experiments on nanotubes grown inside zeolite AFI single crystals. Diffuse scattering intensity in the plane l=0 and at small wave vectors can be interpreted as being due to a partial filling of the zeolite channels by nanotubes whose diameter is about 4Å (upper limit 5Å). This value is compatible with the zeolite channel size. Such nanotubes are among the smallest in diameter reported up to now. It is suggested that interactions with AFI during the nanotube synthesis may lead to the selection of only one type of nanotube.

**Acknowledgements**
The authors thank Alessandra Marucci, Sylvain Ravy, Eric Sandré and Michèle Veber for interesting discussions. They are grateful to the ID01 team at ESRF for the allocated beamtime.

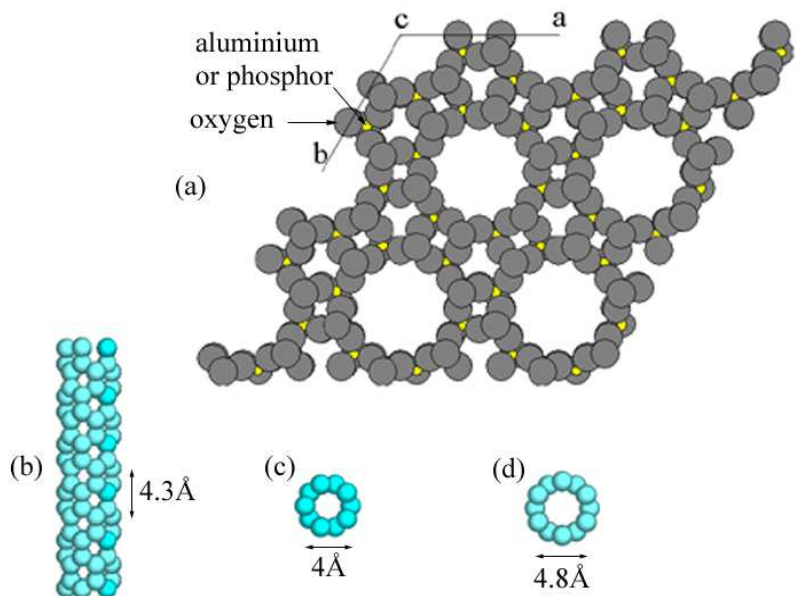

**Figure 1.** (a) Structure of AlPO$_4$-5 viewed along **c** and showing the open channels, (b) view of a (5,0) nanotube perpendicularly to its long axis, (c) (5,0) nanotube projected along its axis, (d) (6,0) nanotube projected along its axis.

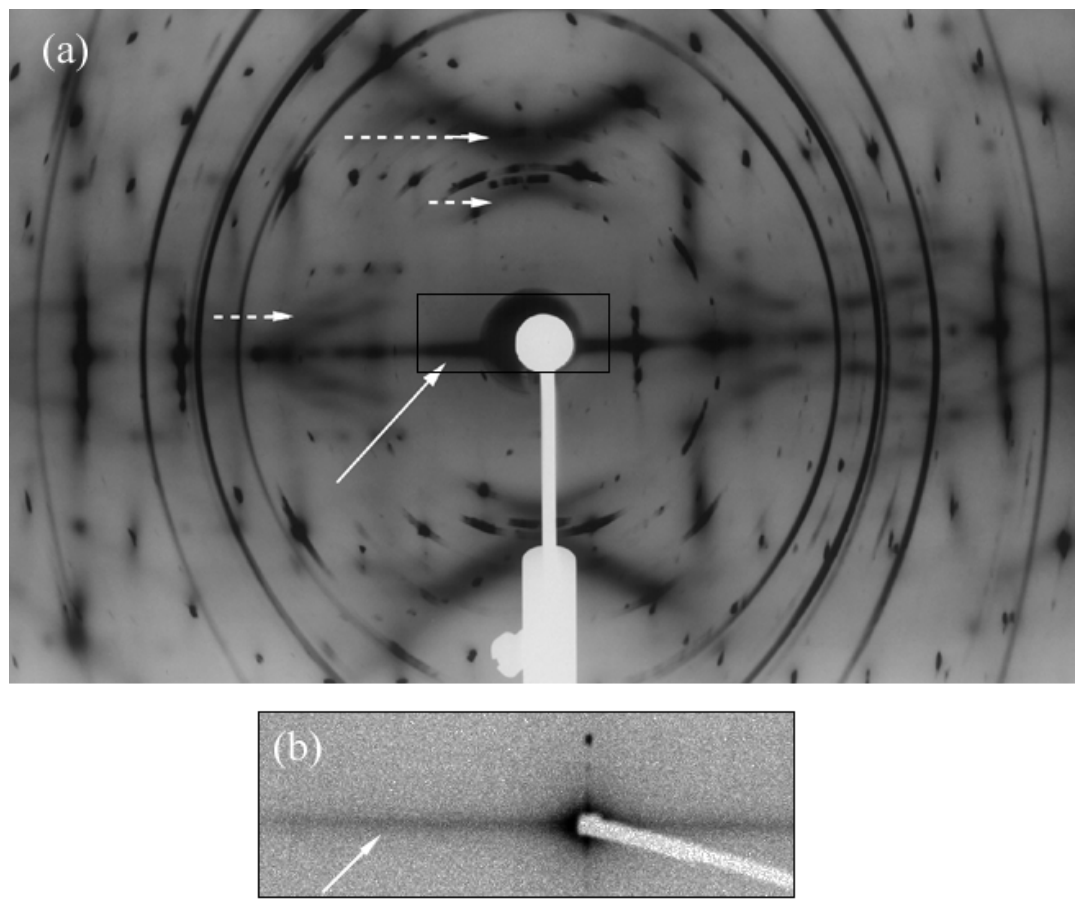

**Figure 2.** Diffraction patterns (crystal-fixed detector-fixed) obtained with an AFI-nanotube crystal, with vertical **c*** axis (a) cylindrical ($\Phi$=57.3mm) film, using an X-ray tube ($\lambda$=1.5418Å), (b) planar gas detector placed at 618mm from the sample, using synchrotron

radiation ($\lambda$=1.044 Å). In (a) and (b), the full arrow points to the l=0 diffuse scattering observed for AFI-nanotube crystals only. In (a), the dotted arrows indicate diffuse scattering features similar to those observed in AFI-hollow or AFI-TPA crystals. Sharp diffraction rings are due to the sample holder. In (a), the first ring at Q≈0.43 Å$^{-1}$ gives the scale of the pattern. The rectangular area in (a) is magnified in (b).

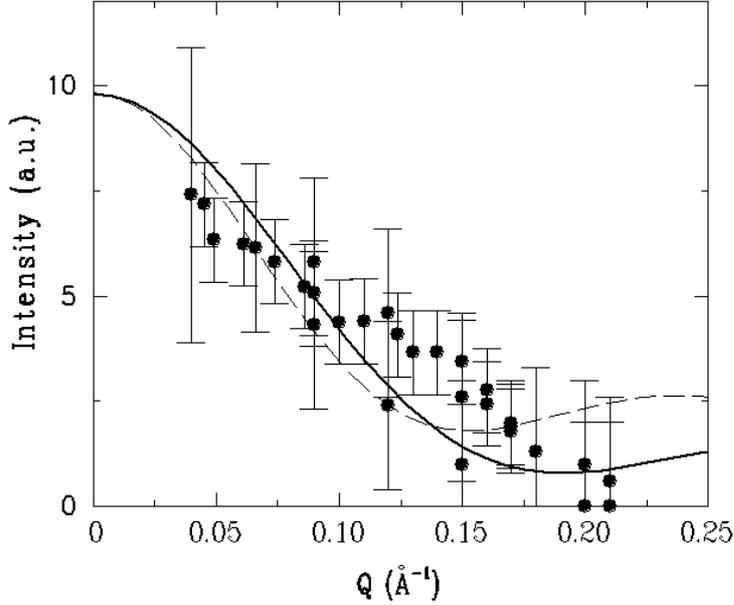

**Figure 3.** Intensity in the plane l=0 as a function of the wave vector Q. Filled circles refer to the experimental data while the solid and dotted lines are calculated for nanotubes diameters of 4 and 5Å, respectively (see text).

| (n,m) | Diameter (Å) | Period (Å) |
|---|---|---|
| (5,0) | 3.97 | 4.32 |
| (3,3) | 4.13 | 2.49 |
| (4,2) | 4.20 | 11.43 |
| (5,1) | 4.42 | 24.05 |
| (6,0) | 4.76 | 4.32 |
| (4,3) | 4.83 | 26.28 |
| (5,2) | 4.96 | 8.99 |

**Table 1.** Nanotube indices (n,m), diameter and period (notation and calculations from ref.3; the indices refer to the 60° basis for the graphene lattice and the diameter is calculated as $\Phi = \frac{\sqrt{3} d_{C-C} \sqrt{n^2 + m^2 + nm}}{\pi}$ with $d_{C-C}$=1.44Å , slightly larger than in graphite.).